\documentclass[12pt,thmsa,notitlepage]{article}
\usepackage{amsmath}
\usepackage{amssymb}
\usepackage{amsfonts}
\usepackage{sw20lart}
\usepackage[onehalfspacing]{setspace}
\usepackage{graphicx}

\input{tcilatex}

\begin{document}

\title{Lorentz Transformation in Flat $5D$ Complex-Hyperbolic Space}
\author{{\large T. Q. Truong\thanks{%
mtruong@fontbonne.edu}} \\
{\small Department of Biological and Physical Sciences }\\
{\small Fontbonne University}\\
St. Louis, MO 63105}
\maketitle

\begin{abstract}
The Lorentz transfomation is derived in $5D$ flat pseudo-complex affine
space or $TT$ Space. The $TT$ space or pseudo-Complex space accomodates one
uncompactified time-like extra dimension. The effects of one extra 
time-like dimension are shown to affect the structure of the Lorentz
transformation and the maximum allowable speed. The maximum allowable speed
for particles living in $TT$ space is shown to exceed the speed of light, $c$%
, where $c$ is the absolute speed in Minkowski space.

\newpage
\end{abstract}

\section{\protect\bigskip Introduction}

Non-Euclidean geometries are geometries with the exclusion of \ Euclid's
fifth postulate, the \textit{parallel postulate\cite{Euclid}}. Minkowski and
Einstein saw the potential application of one of these non-Euclidean
geometries, the hyperbolic geometry. Subsequently, Einstein formulated his
general theory of relativity on the hyperbolic space. The hyperbolic space
with the flat metric $g_{\mu \nu }=$ $diag(-1,1,1,1)=\eta _{\mu \nu },$ is
the well known Minkowski space. The standard approach in dealing with higher
dimensional theories is to consider almost exclusively space-like extra
dimensions. Large extra-dimensions have been used to address the hierarchy
problem, whereas Higgs mass is proven to be finite \cite{Higgsmass}. The
effects of the extra space-like dimensions have also been examined in the
context of 4D superspace formalism \cite{Truong}. However, there is no
priory reason why extra time-like dimensions cannot exist. Time-like extra
dimensions have been ignored due to serious conflicts with causality and
unitarity \cite{Yndurain}\cite{Senjanovic}\cite{Erdem}. However, these
technicalities could be avoided if calculations are performed in $5D$ flat
pseudo-complex affine space with metric function $g_{AB}=\left[ TT\right]
_{AB}=diag(-1,-1,1,1,1)$ or in short, $TT$ space. The acronym, $TT,$ is
associated with the two minuses in the metric function.

The letter is divided as follows. In section 2, the specific forms of
elements, $p^{A}\in M\subset $ $TT$ space and $\overset{\cdot }{p}_{A}\in
W\subset T_{\overrightarrow{p}}(M)$, where $M$ and $W$ are some open subsets
of $TT$ space and its associated tangent space $T_{P^{A}}(M)$, are derived
via $TT$ metric function. In section 3, the $TT$ Lorentz \ transformation is
obtained. In section 4, as an application in $TT$ space, using the $TT$
Lorentz transformation, a boost of $v$ along the spatial direction between
the two frames is perform. The maximum allowable speed for particles living
in $TT$ space is shown to be $v_{\max }<\sqrt{2}c,$ which exceeds the
absolute speed, $c,$ the speed of light in Minkowski space.

\section{$TT$ Space and The Associated Tangent Space}

In this section, elements of $5D$ flat pseudo-complex affine space or $TT$
space with metric $g_{AB}=\left[ TT\right] _{AB}=diag(-1,-1,1,1,1)$ and
Lorentzian signature $\left( r-s\right) =n-2s=1,$ are derived. Consider an
element in our $TT$ space, $p^{A}\in M\subset TT,$ where $M$ is an open set
contained in $TT$ space and $p^{A}$ is a mapping

\begin{eqnarray}
p^{A}\left( \psi \right) &:&%
\mathbb{R}
\rightarrow TT\subset 
\mathbb{C}
^{5}  \notag \\
p^{A}\left( \psi \right) &=&p_{0}+\left( 
\begin{array}{c}
x^{1}\left( \psi \right) \\ 
x^{2}\left( \psi \right) \\ 
x^{3}\left( \psi \right) \\ 
x^{4}\left( \psi \right) \\ 
x^{5}\left( \psi \right)%
\end{array}%
\right)  \notag \\
p^{A}\left( \psi \right) &=&p_{0}+x^{A}\left( \psi \right) ,
\end{eqnarray}%
where $%
\mathbb{R}
$ is the real, $%
\mathbb{C}
^{5}$ is $5D$ complex space, $x^{A}\left( \psi \right) $ are coordinates of
the point $p^{A}\left( \psi \right) ,$ and $p_{0}$ is defined to be the
origin by the initial condition, accompanied by a parametrization parameter $%
\psi $. The range of the mapping function $p^{A}\left( \psi \right) $ can
have non-real values because of the complex nature of the Lorentz
transformations associated with the $TT$ space. The 5-vector can be computed
by taking the total differential of $p^{A}\left( \psi \right) $ 
\begin{eqnarray}
dp^{A}\left( \psi \right) &=&\sum\limits_{B=1}^{5}\frac{\partial p^{A}}{%
\partial x^{B}}dx^{B}  \notag \\
dp^{A}\left( \psi \right) &=&\widehat{x}_{1}dx^{1}\left( \psi \right) +%
\widehat{x}_{2}dx^{2}\left( \psi \right) +\widehat{x}_{3}dx^{3}\left( \psi
\right) +\widehat{x}_{4}dx^{4}\left( \psi \right) +\widehat{x}%
_{5}dx^{5}\left( \psi \right)  \notag \\
dp^{A}\left( \psi \right) &=&\left( 
\begin{array}{c}
dx^{1}\left( \psi \right) \\ 
dx^{2}\left( \psi \right) \\ 
dx^{3}\left( \psi \right) \\ 
dx^{4}\left( \psi \right) \\ 
dx^{5}\left( \psi \right)%
\end{array}%
\right)  \label{5-vector}
\end{eqnarray}%
Without loss of generality, the spatial component of $p^{A}\left( \psi
\right) $ can be suppressed, i.e.%
\begin{equation}
p^{A}\left( \psi \right) =p_{0}+\left( 
\begin{array}{c}
x^{1}\left( \psi \right) \\ 
x^{2}\left( \psi \right) \\ 
r^{i}(\psi )%
\end{array}%
\right) ,
\end{equation}%
and 
\begin{equation}
r^{i}(\psi )=\left( 
\begin{array}{c}
x^{3}\left( \psi \right) \\ 
x^{4}\left( \psi \right) \\ 
x^{5}(\psi )%
\end{array}%
\right) ,
\end{equation}%
where $i=3,4,5.$ The $TT$ space is a subset of $%
\mathbb{C}
^{5},$ and is defined as%
\begin{eqnarray}
TT &=&\left\{ \left( x_{1},x_{2},x_{3},x_{4},x_{5}\right) \in 
\mathbb{C}
^{5}:-x_{1}^{2}-x_{2}^{2}+x_{3}^{2}+x_{4}^{2}+x_{5}^{2}=R^{2}\right\} \\
TT &=&\left\{ \left( x_{1},x_{2},r_{i}\right) \in 
\mathbb{C}
^{5}:-x_{1}^{2}-x_{2}^{2}+r_{i}^{2}=R^{2}\right\} ,
\end{eqnarray}%
where $R$ is the positive constant curvature of $TT$ affine space. The
mapping 
\begin{equation}
p^{A}(\psi )=p_{0}+\left( x^{1}\left( \psi \right) ,x^{2}\left( \psi \right)
,r^{i}\left( \psi \right) \right) ,
\end{equation}%
with the initial condition,%
\begin{equation}
p^{A}(0)=p_{0}+\left( x^{1}\left( 0\right) ,x^{2}\left( 0\right)
,r^{i}\left( 0\right) \right) =\left( 0,0,R\right) ,
\label{Initial Condition}
\end{equation}%
where $r^{i}\left( 0\right) =R,$ would yield $p_{0}=\left( 0,0,0\right) .$
An element of tangent space to $TT$ at $p^{A}(\psi )$ can be obtain by
taking the derivative of $p^{A}(\psi ),$ w.r.t $\psi ,$%
\begin{equation}
\overset{\cdot }{p}_{A}\left( \psi \right) =\frac{dp_{A}}{d\psi }=\left( 
\overset{\cdot }{x}_{1},\overset{\cdot }{x}_{2},\overset{\cdot }{r}%
_{i}\right) ,
\end{equation}%
where $\overset{\cdot }{r}_{i}=\frac{dr_{i}}{d\psi }$and $\overset{\cdot }{x}%
_{j}=\frac{dx_{j}}{d\psi }.$ Taking the $TT$ inner product is defined as $%
\left( \text{ }\circledast \text{ }\right) ,$ and the operation satisfies
the $TT$ metric function%
\begin{eqnarray}
p^{2} &=&p^{A}\circledast p_{A}=g_{AB}p^{A}p^{B}  \label{metric function} \\
p^{2} &=&-x_{1}^{2}-x_{2}^{2}+r_{i}^{2}=R^{2}.
\end{eqnarray}%
Isometries of \ a hyperboloid model associated with ordinary hyperbolic space%
\cite{Cannon}, 
\begin{equation}
L=\left\{ \left( x_{1,}.....,x_{n},x_{n+1}\right) :x_{1}^{2}+\cdot \cdot
\cdot \cdot +x_{n}^{2}-x_{n+1}^{2}\text{ and }x_{n+1}>0\right\} ,
\end{equation}%
where $x_{n+1}$ is the time-like dimension. Transformations or linear maps
of elements from the hyperboloid model which preserve the associated metric
or the hyperbolic inner product%
\begin{equation}
dx_{L}^{2}=dx_{1}^{2}+\cdot \cdot \cdot +dx_{n}^{2}+dx_{n+1}^{2},
\end{equation}%
would induce linear, Riemannian and topological isometries on that space.
Analogously, Lorentz transformations on elements $p^{A}\in M\subset $ $TT$
space, which preserve the $TT$ inner product $\left( \ref{metric function}%
\right) ,$ induces a pseudo-Riemannian isometry $\left[ L\right] $ on $TT.$
The pseudo-Riemannian isometry $\left[ L\right] _{B}^{A}$ is a
diffeomorphism of $TT$ that satisfies 
\begin{equation}
\left[ L\right] _{\ast }(ds^{2})\left( u,v\right) =ds^{2}\left[ D\left[ L%
\right] \left( u\right) ,D\left[ L\right] \left( v\right) \right] ,
\label{pull back}
\end{equation}%
where $\left[ L\right] _{\ast }$ is the pullback, $u$ and $v$ are elements
from the tangent space $T_{P^{A}}(M)$ at point $p^{A}$ and $D\left[ L\right] 
$ is the derivative maps on tangent vectors at $p^{A}$ to tangent vectors at 
$\left[ L\right] \left( p^{A}\right) .$

Differentiating equation $\left( \ref{metric function}\right) $ w.r.t.
parametrization parameter $\psi $, we have 
\begin{equation}
-x_{1}\overset{\cdot }{x}_{1}-x_{2}\overset{\cdot }{x}_{2}+r_{i}\overset{%
\cdot }{r}_{i}=0.  \label{inner product}
\end{equation}%
Equation $\left( \ref{inner product}\right) $ is equivalent to 
\begin{eqnarray}
p^{A}\circledast \overset{\cdot }{p}_{A} &=&\left( x^{1},x^{2},r^{i}\right)
\circledast \left( \overset{\cdot }{x}_{1},\overset{\cdot }{x}_{2},\overset{%
\cdot }{r}_{i}\right) \\
p^{A}\circledast \overset{\cdot }{p}_{A} &=&-x_{1}\overset{\cdot }{x}%
_{1}-x_{2}\overset{\cdot }{x}_{2}+r_{i}\overset{\cdot }{r}_{i}=0.
\end{eqnarray}%
The vectors $p^{A}$ and $\overset{\cdot }{p}_{A}$are said to be $TT$
perpendiculars w.r.t. each other. By inspection, the specific forms of $%
p^{A} $ and $\overset{\cdot }{p}_{A}$can be obtained from the following
differential equations%
\begin{eqnarray}
\overset{\cdot }{r}_{i} &=&x_{1}+x_{2},  \label{1st order dfe 1} \\
\overset{\cdot }{x}_{1} &=&r_{i},  \label{1st order dfe 2} \\
\overset{\cdot }{x}_{2} &=&r_{i}.  \label{1st order dfe 3}
\end{eqnarray}%
Taking the derivative of equation $\left( \ref{1st order dfe 1}\right) $%
yields%
\begin{eqnarray}
\overset{\cdot \cdot }{r}_{i} &=&\overset{\cdot }{x}_{1}+\overset{\cdot }{x}%
_{2}=2r_{i}  \notag \\
\overset{\cdot \cdot }{r}_{i} &=&2r_{i}.  \label{2nd order dfe}
\end{eqnarray}%
The solution to equation $\left( \ref{2nd order dfe}\right) $ takes a
general form of 
\begin{equation}
r_{i}\left( \psi \right) =Ae^{\sqrt{2}\psi }+Be^{-\sqrt{2}\psi },
\label{Sol 1}
\end{equation}%
where $A$ and $B$ are arbitrary constants and to fixed by the initial
condition $\left( \ref{Initial Condition}\right) $. To determine the
constants, take the derivative of equation $\left( \ref{Sol 1}\right) $,
substituting in for equation $\left( \ref{1st order dfe 1}\right) $ 
\begin{equation}
\overset{\cdot }{r}_{i}\left( \psi \right) =\sqrt{2}Ae^{\sqrt{2}\psi }-\sqrt{%
2}Be^{-\sqrt{2}\psi }=x_{1}\left( \psi \right) +x_{2}\left( \psi \right) .
\label{1st order dfe 1.1}
\end{equation}%
Imposing the temporal initial condition $\left( \ref{Initial Condition}%
\right) $ to equation $\left( \ref{1st order dfe 1.1}\right) ,$ yields%
\begin{eqnarray}
\overset{\cdot }{r}_{i}\left( 0\right) &=&\sqrt{2}Ae^{\sqrt{2}\cdot 0}-\sqrt{%
2}Be^{-\sqrt{2}\cdot 0}=x_{1}\left( 0\right) +x_{2}\left( 0\right) =0  \notag
\\
&\Rightarrow &\text{ }A-B=0\text{ \ }\Rightarrow \text{ }A=B.
\end{eqnarray}%
The spatial solution $\left( \ref{Sol 1}\right) $ becomes%
\begin{equation}
r_{i}\left( \psi \right) =A\left( e^{\sqrt{2}\psi }+e^{-\sqrt{2}\psi
}\right) .  \label{Sol 1.1}
\end{equation}%
Imposing the spatial initial condition $\left( \ref{Initial Condition}%
\right) $ on equation $\left( \ref{Sol 1.1}\right) $ 
\begin{eqnarray}
r_{i}\left( 0\right) &=&A\left( e^{\sqrt{2}\cdot 0}+e^{-\sqrt{2}\cdot
0}\right) =R  \notag \\
&\Rightarrow &\text{ }2A=R\text{ }\Rightarrow \text{ }A=\frac{R}{2}.
\end{eqnarray}%
The particular spatial solution is 
\begin{eqnarray}
r_{i}\left( \psi \right) &=&R\left( \frac{e^{\sqrt{2}\psi }+e^{-\sqrt{2}\psi
}}{2}\right)  \notag \\
r_{i}\left( \psi \right) &=&R\cosh \left( \sqrt{2}\psi \right) .
\label{Sol 1.1.1}
\end{eqnarray}%
Using the spatial solution $\left( \ref{Sol 1.1.1}\right) ,$ the two
temporal solutions for equations $\left( \ref{1st order dfe 2}\right) $ and$%
\left( \ref{1st order dfe 3}\right) $can now be obtained%
\begin{eqnarray}
\frac{dx_{1}\left( \psi \right) }{d\psi } &=&R\cosh \left( \sqrt{2}\psi
\right) \text{ \ \ }\Rightarrow \text{ \ \ }x_{1}\left( \psi \right) =\frac{R%
}{\sqrt{2}}\sinh \left( \sqrt{2}\psi \right) +C_{1},  \label{Sol 2} \\
\frac{dx_{2}\left( \psi \right) }{d\psi } &=&R\cosh \left( \sqrt{2}\psi
\right) \text{ \ \ }\Rightarrow \text{ \ \ }x_{2}\left( \psi \right) =\frac{R%
}{\sqrt{2}}\sinh \left( \sqrt{2}\psi \right) +C_{2}.  \label{Sol 3}
\end{eqnarray}%
Imposing the temporal initial condition $\left( \ref{Initial Condition}%
\right) $ on solutions $\left( \ref{Sol 2}\right) $ and $\left( \ref{Sol 3}%
\right) ,$ yields vanishing constants $C_{1}$ and $C_{2},$ 
\begin{eqnarray}
\text{ \ \ }x_{1}\left( \psi \right) &=&\frac{R}{\sqrt{2}}\sinh \left( \sqrt{%
2}\psi \right) ,  \label{Sol 2.1} \\
\text{ \ \ }x_{2}\left( \psi \right) &=&\frac{R}{\sqrt{2}}\sinh \left( \sqrt{%
2}\psi \right) .  \label{Sol 3.1}
\end{eqnarray}%
Parameterized by $\psi ,$ the temporal and spatial components of \ $p^{A}\in
M\subset $ $TT$ can be specified as 
\begin{equation}
p^{\widehat{A}}\left( \psi \right) =\frac{R}{\sqrt{2}}\left( \sinh \left( 
\sqrt{2}\psi \right) ,\sinh \left( \sqrt{2}\psi \right) ,\sqrt{2}\cosh
\left( \sqrt{2}\psi \right) \right) ,  \label{element of TT space}
\end{equation}%
where $\widehat{A}=1,2,3.$ The $TT$ perpendicular to $p^{\widehat{A}}$ is%
\begin{equation}
\overset{\cdot }{p}_{\widehat{A}}\left( \psi \right) =R\left( \cosh \left( 
\sqrt{2}\psi \right) ,\cosh \left( \sqrt{2}\psi \right) ,\sqrt{2}\sinh
\left( \sqrt{2}\psi \right) \right) .
\end{equation}%
The following are additional relations between $TT$ space and its associated
tangent space $T_{p^{\widehat{A}}}(M),$%
\begin{eqnarray}
p^{2} &=&R^{2} \\
\overset{\cdot }{p}^{2} &=&-2R^{2}=-p\circledast \overset{\cdot \cdot }{p},
\end{eqnarray}%
where 
\begin{equation}
\overset{\cdot \cdot }{p}_{\widehat{A}}\left( \psi \right) =R\sqrt{2}\left(
\sinh \left( \sqrt{2}\psi \right) ,\sinh \left( \sqrt{2}\psi \right) ,\sqrt{2%
}\cosh \left( \sqrt{2}\psi \right) \right)
\end{equation}%
the second derivative of $p^{\widehat{A}}\left( \psi \right) .$

\section{Lorentz Transformation in $TT$ Space}

In this section, the Lorentz transformation in $TT$ space is derived. Recall
from section 2, that an element of $TT$ space is parametrized by a single
parameter $\psi $%
\begin{eqnarray}
p^{\widehat{A}}\left( \psi \right) &=&\frac{R}{\sqrt{2}}\left( \sinh \left( 
\sqrt{2}\psi \right) ,\sinh \left( \sqrt{2}\psi \right) ,\sqrt{2}\cosh
\left( \sqrt{2}\psi \right) \right) , \\
p^{A}\left( \psi \right) &=&\left( ct_{1}\left( \psi \right) ,ct_{2}\left(
\psi \right) ,r_{i}\left( \psi \right) \right) ,
\end{eqnarray}%
where $t_{1}$, $t_{2}$, $r_{i}$ are the two time-like and three space-like
dimensions and $\psi \in 
\mathbb{R}
.$ The $p^{\widehat{A}}\left( \psi \right) $ and $p^{A}\left( \psi \right) $
are 3-D and 5-D vectors, respectively. The total differential of $p^{A}$ is 
\begin{equation*}
dp^{A}\left( \psi \right) =\left( cdt_{1}\left( \psi \right) ,cdt_{2}\left(
\psi \right) ,dr_{i}\left( \psi \right) \right) .
\end{equation*}%
The Lorentz transformation in $TT$ space can be written in matrix equation
as 
\begin{equation}
dp^{\prime A}=\left[ L\right] _{B}^{A}dp^{B},
\end{equation}%
where $dp^{B}$ is a 5-vector and $\left[ L\right] _{B}^{A}$ is the $5\times
5 $ Lorentz transformation matrix. Elements of $\left[ L\right] _{B}^{A}$
can be fixed by length-invariance condition%
\begin{equation}
\left[ L\right] \left[ TT\right] \left[ \overline{L}\right] ^{T}=\left[ TT%
\right] ,  \label{Length invariance condition}
\end{equation}%
where 
\begin{equation*}
\left[ TT\right] =\left( 
\begin{array}{ccccc}
-1 & 0 & 0 & 0 & 0 \\ 
0 & -1 & 0 & 0 & 0 \\ 
0 & 0 & 1 & 0 & 0 \\ 
0 & 0 & 0 & 1 & 0 \\ 
0 & 0 & 0 & 0 & 1%
\end{array}%
\right) ,
\end{equation*}%
and $\left[ \overline{L}\right] ^{T}$ is the complex conjugate-transposed
matrix. To obtain the rotation matrix on the $r_{4}-r_{5}$ plane, we rotate
around the $r_{3}$ by an angle $\theta .$ The rotation transformation 
\begin{equation}
\left( 
\begin{array}{c}
dx_{4}^{\prime } \\ 
dx_{5}^{\prime }%
\end{array}%
\right) =R(\theta )_{r_{4}-r_{5}}\left( 
\begin{array}{c}
dx_{4} \\ 
dx_{5}%
\end{array}%
\right) ,
\end{equation}%
with%
\begin{equation}
R(\theta )_{r_{4}-r_{5}}=\left( 
\begin{array}{cc}
\cos \theta & -\sin \theta \\ 
\sin \theta & \cos \theta%
\end{array}%
\right) ,
\end{equation}%
preserves the length invariance of the 5-vector $dp^{A}$. Likewise, rotation
transformation on the $ct_{1}-ct_{2}$ plane which preserves length
invariance of the 5-vector $dp^{A}$ can be performed by%
\begin{equation}
\left( 
\begin{array}{c}
cdt_{1}^{\prime } \\ 
cdt_{2}^{\prime }%
\end{array}%
\right) =R(\phi )_{ct_{1}-ct_{2}}\left( 
\begin{array}{c}
cdt_{1} \\ 
cdt_{2}%
\end{array}%
\right) ,
\end{equation}%
where 
\begin{equation}
R(\phi )_{ct_{1}-ct_{2}}=\left( 
\begin{array}{cc}
-\cos \phi & +\sin \phi \\ 
-\sin \phi & -\cos \phi%
\end{array}%
\right) .
\end{equation}%
The boost transformation between two inertial frames by $\overrightarrow{v}=v%
\widehat{r}_{3}$, where $\widehat{r}_{3}$ is a unit vector pointing in the $%
r_{3}$-direction. With this boost, $\left( r_{4},r_{5}\right) $ coordinates
are not affected, hence we could rewrite the 5-vector $dp^{A}$ as 3-vector $%
dp^{\widehat{A}}$%
\begin{equation}
dp^{\widehat{A}}=\left( cdt_{1}\left( \psi \right) ,cdt_{2}\left( \psi
\right) ,dr_{3}\left( \psi \right) \right) ,  \label{3-vector}
\end{equation}%
where $\widehat{A}=1,2,3.$ The transformed 3-vector can be obtained by 
\begin{eqnarray}
dp^{^{\prime }\widehat{A}} &=&\left[ L\right] _{\widehat{B}}^{\widehat{A}%
}dp^{\widehat{B}}  \label{TT Lorentz transformation} \\
\left( 
\begin{array}{c}
ct_{1}^{\prime } \\ 
ct_{2}^{\prime } \\ 
r_{3}^{\prime }%
\end{array}%
\right) &=&\left( 
\begin{array}{ccc}
a & b & c \\ 
d & e & f \\ 
g & h & i%
\end{array}%
\right) \left( 
\begin{array}{c}
ct_{1} \\ 
ct_{2} \\ 
r_{3}%
\end{array}%
\right) ,  \notag
\end{eqnarray}%
where $\widehat{B}=1,2,3.$ The length invariance condition requires%
\begin{eqnarray}
\left( 
\begin{array}{ccc}
a & b & c \\ 
d & e & f \\ 
g & h & i%
\end{array}%
\right) \left( 
\begin{array}{ccc}
-1 & 0 & 0 \\ 
0 & -1 & 0 \\ 
0 & 0 & 1%
\end{array}%
\right) \left( 
\begin{array}{ccc}
a & d & g \\ 
b & e & h \\ 
c & f & i%
\end{array}%
\right) &=&\left( 
\begin{array}{ccc}
-1 & 0 & 0 \\ 
0 & -1 & 0 \\ 
0 & 0 & 1%
\end{array}%
\right)  \notag \\
\left( 
\begin{array}{ccc}
a & b & c \\ 
d & e & f \\ 
g & h & i%
\end{array}%
\right) \left( 
\begin{array}{ccc}
-a & -d & -g \\ 
-b & -e & -h \\ 
c & f & i%
\end{array}%
\right) &=&\left( 
\begin{array}{ccc}
-1 & 0 & 0 \\ 
0 & -1 & 0 \\ 
0 & 0 & 1%
\end{array}%
\right)  \notag \\
\left( 
\begin{array}{ccc}
-\left( a^{2}+b^{2}\right) +c^{2} & -ad-be+cf & -ag-bh+ci \\ 
-da-eb+fc & -(d^{2}+e^{2})+f^{2} & -dg-eh+fi \\ 
-ga-hb+ic & -gd-he+if & -(g^{2}+h^{2})+i^{2}%
\end{array}%
\right) &=&\left( 
\begin{array}{ccc}
-1 & 0 & 0 \\ 
0 & -1 & 0 \\ 
0 & 0 & 1%
\end{array}%
\right) .
\end{eqnarray}%
The length invariance condition yields the following equations%
\begin{eqnarray}
-\left( a^{2}+b^{2}\right) +c^{2} &=&-1,  \label{Constraint 1} \\
-(d^{2}+e^{2})+f^{2} &=&-1,  \label{Constraint 2} \\
-(g^{2}+h^{2})+i^{2} &=&1,  \label{Constraint 3}
\end{eqnarray}%
and 
\begin{equation}
-ad-be+cf=-ag-bh+ci=-dg-eh+fi=0,  \label{Constraint 4-6}
\end{equation}%
which can then be used to fix the boost transformation matrix $\left[ L%
\right] _{\widehat{B}}^{\widehat{A}}.$ Using equations $\left( \ref%
{Constraint 1}\right) $ to $\left( \ref{Constraint 3}\right) ,$ the matrix
becomes 
\begin{eqnarray}
\left[ L\right] _{\widehat{B}}^{\widehat{A}} &=&\left( 
\begin{array}{ccc}
a & b & c \\ 
d & e & f \\ 
g & h & i%
\end{array}%
\right)  \notag \\
&=&\left( 
\begin{array}{ccc}
\frac{1}{\sqrt{2}}\cosh \left( \sqrt{2}\psi _{1}\right) & \frac{1}{\sqrt{2}}%
\cosh \left( \sqrt{2}\psi _{1}\right) & \sinh \left( \sqrt{2}\psi _{1}\right)
\\ 
\frac{1}{\sqrt{2}}\cosh \left( \sqrt{2}\psi _{2}\right) & \frac{1}{\sqrt{2}}%
\cosh \left( \sqrt{2}\psi _{2}\right) & \sinh \left( \sqrt{2}\psi _{2}\right)
\\ 
\frac{1}{\sqrt{2}}\sinh \left( \sqrt{2}\psi _{3}\right) & \frac{1}{\sqrt{2}}%
\sinh \left( \sqrt{2}\psi _{3}\right) & \cosh \left( \sqrt{2}\psi _{3}\right)%
\end{array}%
\right) ,
\end{eqnarray}%
where $\psi _{j}\in 
\mathbb{R}
$ are parametrization parameters, $\ j=1,2,3$. The relationships among the
parametrization parameters can be obtained through $\left( \ref{Constraint
4-6}\right) ,$%
\begin{eqnarray}
-ad-be+cf &=&0  \notag \\
-\cosh \left( \sqrt{2}\psi _{1}\right) \cosh \left( \sqrt{2}\psi _{2}\right)
+\sinh \left( \sqrt{2}\psi _{1}\right) \cosh \left( \sqrt{2}\psi _{2}\right)
&=&0  \notag \\
\cosh \left[ \sqrt{2}\left( \psi _{1}-\psi _{2}\right) \right] &=&0.
\label{relation 1}
\end{eqnarray}%
Equation $\left( \ref{relation 1}\right) $ can be satisfied if $\sqrt{2}%
\left( \psi _{1}-\psi _{2}\right) =i\pi \left( n+\frac{1}{2}\right) ,$ thus
we must have%
\begin{equation}
\psi _{2}=\psi _{1}-i\frac{\pi }{\sqrt{2}}\left( n+\frac{1}{2}\right) .
\end{equation}%
The relation between $\psi _{1}$ and $\psi _{3}$ is obtained by $\left( \ref%
{Constraint 4-6}\right) $ 
\begin{eqnarray}
-ag-bh+ci &=&0  \notag \\
-\cosh \left( \sqrt{2}\psi _{1}\right) \sinh \left( \sqrt{2}\psi _{3}\right)
+\sinh \left( \sqrt{2}\psi _{1}\right) \cosh \left( \sqrt{2}\psi _{3}\right)
&=&0  \notag \\
\sinh \left[ \sqrt{2}\left( \psi _{1}-\psi _{3}\right) \right] &=&0,
\label{relation 2}
\end{eqnarray}%
implies $\psi _{3}=\psi _{1}.$ The boost transformation matrix becomes%
\begin{equation}
\left[ L\right] _{\widehat{B}}^{\widehat{A}}=\left( 
\begin{array}{ccc}
\frac{1}{\sqrt{2}}\cosh \left( \sqrt{2}\psi _{1}\right) & \frac{1}{\sqrt{2}}%
\cosh \left( \sqrt{2}\psi _{1}\right) & \sinh \left( \sqrt{2}\psi _{1}\right)
\\ 
\frac{1}{\sqrt{2}}\cosh \sqrt{2}z & \frac{1}{\sqrt{2}}\cosh \sqrt{2}z & 
\sinh \left[ \sqrt{2}z\right] \\ 
\frac{1}{\sqrt{2}}\sinh \left( \sqrt{2}\psi _{1}\right) & \frac{1}{\sqrt{2}}%
\sinh \left( \sqrt{2}\psi _{1}\right) & \cosh \left( \sqrt{2}\psi _{1}\right)%
\end{array}%
\right) ,
\end{equation}%
where $z=\left( \psi _{1}-i\frac{\pi }{\sqrt{2}}\left( n+\frac{1}{2}\right)
\right) .$ After little algebra, the Lorentz transformation matrix $\left[ L%
\right] _{\widehat{B}}^{\widehat{A}}$ simplifies to%
\begin{equation}
\left[ L\right] _{\widehat{B}}^{\widehat{A}}=\left( 
\begin{array}{ccc}
\frac{1}{\sqrt{2}}C_{1} & \frac{1}{\sqrt{2}}C_{1} & S_{1} \\ 
i\frac{\left( -1\right) ^{n+1}}{\sqrt{2}}S_{1} & i\frac{\left( -1\right)
^{n+1}}{\sqrt{2}}S_{1} & i\left( -1\right) ^{n+1}C_{1} \\ 
\frac{1}{\sqrt{2}}S_{1} & \frac{1}{\sqrt{2}}S_{1} & C_{1}%
\end{array}%
\right) ,
\end{equation}%
where 
\begin{equation}
C_{1}=\cosh \left( \sqrt{2}\psi _{1}\right) ,  \label{substitution 1}
\end{equation}%
and 
\begin{equation}
S_{1}=\sinh \left( \sqrt{2}\psi _{1}\right) .  \label{substitution 2}
\end{equation}%
The Lorentz transformation equation for a 3-vector in $TT$ space becomes 
\begin{eqnarray}
dp^{^{\prime }\widehat{A}} &=&\left[ L\right] _{\widehat{B}}^{\widehat{A}%
}dp^{\widehat{B}}  \notag \\
\left( 
\begin{array}{c}
ct_{1}^{\prime } \\ 
ct_{2}^{\prime } \\ 
r_{3}^{\prime }%
\end{array}%
\right) &=&\left( 
\begin{array}{ccc}
\frac{1}{\sqrt{2}}C_{1} & \frac{1}{\sqrt{2}}C_{1} & S_{1} \\ 
i\frac{\left( -1\right) ^{n+1}}{\sqrt{2}}S_{1} & i\frac{\left( -1\right)
^{n+1}}{\sqrt{2}}S_{1} & i\left( -1\right) ^{n+1}C_{1} \\ 
\frac{1}{\sqrt{2}}S_{1} & \frac{1}{\sqrt{2}}S_{1} & C_{1}%
\end{array}%
\right) \left( 
\begin{array}{c}
ct_{1} \\ 
ct_{2} \\ 
r_{3}%
\end{array}%
\right) ,  \label{TT Lorentz transformation 1}
\end{eqnarray}%
where $i=\sqrt{-1}$ is an imaginary number. Lets define $ct^{\prime
}=c\left( t_{1}^{\prime }+t_{2}^{\prime }\right) $ and $ct=c\left(
t_{1}+t_{2}\right) $, equation $\left( \ref{TT Lorentz transformation 1}%
\right) $ reduces to 
\begin{eqnarray}
\left( 
\begin{array}{c}
ct^{\prime } \\ 
r_{3}^{\prime }%
\end{array}%
\right) &=&\left( 
\begin{array}{cc}
\frac{1}{\sqrt{2}}\left[ C_{_{1}}+i\left( -1\right) ^{n+1}S_{_{1}}\right] & 
S_{_{1}}+i\left( -1\right) ^{n+1}C_{_{1}} \\ 
\frac{1}{\sqrt{2}}S_{_{1}} & C_{_{1}}%
\end{array}%
\right) \left( 
\begin{array}{c}
ct \\ 
r_{3}%
\end{array}%
\right) ,  \notag \\
\left( 
\begin{array}{c}
ct^{\prime } \\ 
r_{3}^{\prime }%
\end{array}%
\right) &=&\left( 
\begin{array}{cc}
\frac{1}{\sqrt{2}}z_{1} & z_{2} \\ 
\frac{1}{\sqrt{2}}S_{1} & C_{1}%
\end{array}%
\right) \left( 
\begin{array}{c}
ct \\ 
r_{3}%
\end{array}%
\right) ,  \label{TT Lorentz transformation 1.1}
\end{eqnarray}%
where 
\begin{equation}
z_{1}=C_{_{1}}+i\left( -1\right) ^{n+1}S_{_{1}},  \label{substitution 3}
\end{equation}%
and%
\begin{equation*}
z_{2}=S_{_{1}}+i\left( -1\right) ^{n+1}C_{_{1}}.
\end{equation*}%
Recall equation $\left( \ref{TT Lorentz transformation 1.1}\right) ,$%
\begin{eqnarray*}
\left( 
\begin{array}{c}
ct^{\prime } \\ 
r_{3}^{\prime }%
\end{array}%
\right) &=&\left( 
\begin{array}{cc}
\frac{1}{\sqrt{2}}\left[ C_{1}+i\left( -1\right) ^{n+1}S_{1}\right] & 
S_{1}+i\left( -1\right) ^{n+1}C_{1} \\ 
\frac{1}{\sqrt{2}}S_{1} & C_{1}%
\end{array}%
\right) \left( 
\begin{array}{c}
ct \\ 
r_{3}%
\end{array}%
\right) \\
\left( 
\begin{array}{c}
ct^{\prime } \\ 
r_{3}^{\prime }%
\end{array}%
\right) &=&C_{1}\left( 
\begin{array}{cc}
\frac{1}{\sqrt{2}}\left[ 1+i\left( -1\right) ^{n+1}T_{1}\right] & 
T_{1}+i\left( -1\right) ^{n+1} \\ 
\frac{1}{\sqrt{2}}T_{1} & 1%
\end{array}%
\right) \left( 
\begin{array}{c}
ct \\ 
r_{3}%
\end{array}%
\right) \\
\left( 
\begin{array}{c}
ct^{\prime } \\ 
r_{3}^{\prime }%
\end{array}%
\right) &=&\gamma _{5}\left( 
\begin{array}{cc}
\frac{1}{\sqrt{2}}\left[ 1-i\frac{\left( -1\right) ^{n+1}}{\sqrt{2}}\beta %
\right] & -\frac{1}{\sqrt{2}}\beta +i\left( -1\right) ^{n+1} \\ 
-\frac{1}{2}\beta & 1%
\end{array}%
\right) \left( 
\begin{array}{c}
ct \\ 
r_{3}%
\end{array}%
\right) \\
\left( 
\begin{array}{c}
ct^{\prime } \\ 
r_{3}^{\prime }%
\end{array}%
\right) &=&\gamma _{5}\left( 
\begin{array}{cc}
\frac{1}{\sqrt{2}}\left[ 1-i\left( -1\right) ^{n+1}\beta _{5}\right] & 
-\beta _{5}+i\left( -1\right) ^{n+1} \\ 
-\frac{1}{\sqrt{2}}\beta _{5} & 1%
\end{array}%
\right) \left( 
\begin{array}{c}
ct \\ 
r_{3}%
\end{array}%
\right) ,
\end{eqnarray*}%
where $T_{1}=\tanh \left( \sqrt{2}\psi _{1}\right) =-\frac{1}{\sqrt{2}}\beta
=-\beta _{5}$ and $C_{1}=\cosh \left( \sqrt{2}\psi _{1}\right) =\gamma _{5}.$
To Simplify the Lorentz transformation matrix, we make the following
substitutions%
\begin{eqnarray}
z_{1,n} &=&\frac{\gamma _{5}}{\sqrt{2}}\left[ 1-i\left( -1\right)
^{n+1}\beta _{5}\right] ,  \label{substitution 5} \\
z_{2,m} &=&\gamma _{5}\left[ 1+i\left( -1\right) ^{-\left( m+1\right) }\beta
_{5}\right] .  \label{substitution 6}
\end{eqnarray}%
The operators $z_{1,n}$ and $z_{2,m}$ can be decomposed into even and odd
parts%
\begin{eqnarray}
z_{1,n} &=&z_{1,n=2k}+z_{1,n=2k+1}  \label{operator 1} \\
z_{1,n} &=&z_{1,even}+z_{1,odd},  \notag
\end{eqnarray}%
and%
\begin{eqnarray}
z_{2,m} &=&z_{2,m=2l}+z_{2,n=2l+1}  \label{operator 2} \\
z_{2,m} &=&z_{2,even}+z_{2,odd},  \notag
\end{eqnarray}%
where $k,l\in 
\mathbb{R}
.$ By inspection, we have the followings%
\begin{eqnarray}
z_{1,n=2k} &=&\frac{\gamma _{5}}{\sqrt{2}}\left[ 1+i\beta _{5}\right] , \\
z_{1,n=2k+1} &=&\frac{\gamma _{5}}{\sqrt{2}}\left[ 1-i\beta _{5}\right] , \\
z_{2,m=2l} &=&\gamma _{5}\left[ 1-i\beta _{5}\right] , \\
z_{2,m=2l+1} &=&\gamma _{5}\left[ 1+i\beta _{5}\right] ,
\end{eqnarray}%
thus we have%
\begin{equation}
z_{2,m=2l}=\sqrt{2}z_{1,n=2k+1},
\end{equation}%
and 
\begin{equation}
z_{2,m=2l+1}=\sqrt{2}z_{1,n=2k}.
\end{equation}%
Equations $\left( \ref{operator 1}\right) $ and $\left( \ref{operator 2}%
\right) $ become%
\begin{eqnarray}
z_{1,n} &=&z_{1,even}+z_{1,odd}=z_{1,even}+\frac{1}{\sqrt{2}}z_{2,even}
\label{operator 3} \\
z_{2,m} &=&z_{2,even}+z_{2,odd}=z_{2,even}+\sqrt{2}z_{1,even}.
\label{operator 4}
\end{eqnarray}%
Multiplying equation $\left( \ref{operator 3}\right) $ by $\sqrt{2}$ and
then divide $\sqrt{2}\left( \ref{operator 3}\right) $ by equation $\left( %
\ref{operator 4}\right) ,$ yields the following relation%
\begin{equation}
z_{2,m}=\sqrt{2}z_{1,n}.  \label{relation 3}
\end{equation}%
The Lorentz transformation matrix then simplifies to%
\begin{equation*}
\left[ L\right] _{\widehat{B}}^{\widehat{A}}=\left( 
\begin{array}{cc}
z_{1,n} & \sqrt{2}z_{1,n} \\ 
-\frac{\gamma _{5}\beta _{5}}{\sqrt{2}} & \gamma _{5}%
\end{array}%
\right) .
\end{equation*}%
The coordinate transformation matrix equation for a 3-vector $dp^{\widehat{A}%
}$ becomes%
\begin{eqnarray}
dp^{^{\prime }\widehat{A}} &=&\left[ L\right] _{\widehat{B}}^{\widehat{A}%
}dp^{\widehat{B}}  \notag \\
\left( 
\begin{array}{c}
cdt^{\prime } \\ 
dr_{3}^{\prime }%
\end{array}%
\right) &=&\left( 
\begin{array}{cc}
z_{1,n} & \sqrt{2}z_{1,n} \\ 
-\frac{\gamma _{5}\beta _{5}}{\sqrt{2}} & \gamma _{5}%
\end{array}%
\right) \left( 
\begin{array}{c}
cdt \\ 
dr_{3}%
\end{array}%
\right)  \notag \\
\left( 
\begin{array}{c}
cdt^{\prime } \\ 
dr_{3}^{\prime }%
\end{array}%
\right) &=&\left( 
\begin{array}{cc}
\frac{\gamma _{5}}{\sqrt{2}}\left[ 1-i\left( -1\right) ^{n+1}\beta _{5}%
\right] & \gamma _{5}\left[ 1-i\left( -1\right) ^{n+1}\beta _{5}\right] \\ 
-\frac{\gamma _{5}\beta _{5}}{\sqrt{2}} & \gamma _{5}%
\end{array}%
\right) \left( 
\begin{array}{c}
ct \\ 
r_{3}%
\end{array}%
\right) ,  \label{TT Lorentz transformation 1.1.1}
\end{eqnarray}%
with the aid of equations $\left( \ref{substitution 5}\right) $ and $\left( %
\ref{relation 3}\right) .$ The inverse $TT$ transformation to $\left[ L%
\right] _{\widehat{B}}^{\widehat{A}}$ is obtained via the length invariant
condition $\left( \ref{Length invariance condition}\right) $ and the
following substitution%
\begin{equation}
z_{1,n}=\frac{\gamma _{5}}{\sqrt{2}}\left[ 1-i\left( -1\right) ^{n+1}\beta
_{5}\right] =\frac{\gamma _{5}}{\sqrt{2}}\xi _{n}(\beta _{5}).
\label{substitution 7}
\end{equation}%
With equation $\left( \ref{substitution 7}\right) $, the $TT$ transformation
matrix $\left( \ref{TT Lorentz transformation 1.1.1}\right) $ reduces to 
\begin{equation*}
\left[ L\right] _{\widehat{B}}^{\widehat{A}}=z_{1,n}\left( 
\begin{array}{cc}
1 & \sqrt{2} \\ 
-\frac{\beta _{5}}{\xi _{n}(\beta _{5})} & \frac{\sqrt{2}}{\xi _{n}(\beta
_{5})}%
\end{array}%
\right) ,
\end{equation*}%
where $\xi _{n}(\beta _{5})=1-i\left( -1\right) ^{n+1}\beta _{5}$ is a
complex-valued function. Without loss of generality, the indices of the $TT$
transformation matrix are suppressed for clarity. Recall equation $\left( %
\ref{Length invariance condition}\right) ,$ we have%
\begin{equation}
\left[ L\right] I_{L}\left[ \overline{L}\right] ^{T}=I_{L},
\label{Length invariance condition 1}
\end{equation}%
where the $TT$ metric matrix 
\begin{equation}
I_{L}=\left( 
\begin{array}{cc}
-1 & 0 \\ 
0 & 1%
\end{array}%
\right) ,  \label{TT metric matrix}
\end{equation}%
and $\left[ \overline{L}\right] ^{T}$ is the complex conjugate-transposed of 
$\left[ L\right] .$ Multiplying both sides of equation $\left( \ref{Length
invariance condition 1}\right) $ by the $\left[ TT\right] $ metric or its $%
2D $ analogue $I_{L}$, we then have 
\begin{equation}
\left[ L\right] I_{L}\left[ \overline{L}\right] ^{T}I_{L}=I_{L}I_{L}=I,
\label{Length invariance condition 2}
\end{equation}%
where 
\begin{equation*}
I_{L}=\left( 
\begin{array}{cc}
-1 & 0 \\ 
0 & 1%
\end{array}%
\right) ,
\end{equation*}%
and 
\begin{equation*}
I=\left( 
\begin{array}{cc}
-1 & 0 \\ 
0 & 1%
\end{array}%
\right) .
\end{equation*}%
Lets define $\left[ L\right] I_{L}=\widehat{\left[ L\right] },$ equation $%
\left( \ref{Length invariance condition 2}\right) $ yields 
\begin{equation}
\widehat{\left[ L\right] }\widehat{\left[ L\right] }^{-1}=I.
\label{Length invariance condition 1.1}
\end{equation}%
The inverse $\widehat{\left[ L\right] }^{-1}$ of $\ \widehat{\left[ L\right] 
}$ can be obtained by standard methods and its closed analytical form yields 
\begin{equation*}
\widehat{\left[ L\right] }^{-1}=\frac{1}{z_{1,n}\left( 1+\beta _{5}\right) }%
\left( 
\begin{array}{cc}
-1 & \xi _{n}(\beta _{5}) \\ 
\frac{1}{\xi _{n}(\beta _{5})}\beta _{5} & \frac{1}{\sqrt{2}}\xi _{n}(\beta
_{5})%
\end{array}%
\right) ,
\end{equation*}%
where $\beta _{5}=\frac{1}{\sqrt{2}}\beta .$ Using equations $\left( \ref%
{Length invariance condition 2}\right) $ and $\left( \ref{Length invariance
condition 1.1}\right) $, the inverse matrix $_{\widehat{B}}^{\widehat{A}}%
\left[ L\right] ^{-1},$ of the $TT$ transformation matrix $\left[ L\right] _{%
\widehat{B}}^{\widehat{A}}$ can finally be obtained%
\begin{eqnarray}
\widehat{\left[ L\right] }\widehat{\left[ L\right] }^{-1} &=&I  \notag \\
\left[ L\right] I_{L}\widehat{\left[ L\right] }^{-1} &=&I
\label{Length invariance condition 3} \\
\left[ L\right] I_{L}\widehat{\left[ L\right] }^{-1}I_{L} &=&II_{L}=I_{L}.
\end{eqnarray}%
Therefore, the inverse $TT$ Lorentz transformation matrix yields 
\begin{eqnarray}
_{\widehat{B}}^{\widehat{A}}\left[ L\right] ^{-1} &=&I_{L}\widehat{\left[ L%
\right] }^{-1}I_{L}  \notag \\
_{\widehat{B}}^{\widehat{A}}\left[ L\right] ^{-1} &=&\frac{1}{z_{1,n}\left(
1+\beta _{5}\right) }\left( 
\begin{array}{cc}
-1 & -\xi _{n}(\beta _{5}) \\ 
-\frac{1}{\sqrt{2}}\beta _{5} & \frac{1}{\sqrt{2}}\xi _{n}(\beta _{5})%
\end{array}%
\right) .
\end{eqnarray}%
The pseudo-Riemannian isometry properties on $TT$ space is evident when we
perform the Lorentz transformations on the infinitesimal length squared 
\begin{eqnarray*}
dp\circledast dp\text{ \ \ }\overset{LT}{\Longrightarrow }\text{ \ }%
dp^{\prime }\circledast dp^{\prime } &=&\text{ }dp^{\widehat{C}}\left( _{%
\widehat{C}}^{\widehat{A}}\left[ L\right] ^{-1}\right) \cdot \left( \left[ L%
\right] _{\widehat{A}}^{\widehat{B}}\right) dp_{\widehat{B}} \\
dp^{\prime }\circledast dp^{\prime } &=&dp^{\widehat{C}}g_{\widehat{C}}^{%
\widehat{B}}dp_{\widehat{B}}=dp\circledast dp \\
dp^{\prime }\circledast dp^{\prime } &=&dp\circledast dp,
\end{eqnarray*}%
which yield an invariant quantity under coordinate transformations.

\bigskip As an application in $TT$ space, the maximum allowable speed for
particles living in $TT$ space is derived. Recall the boost transformation
of parameter $v,$ along the $r_{3}$ direction%
\begin{equation}
r_{3}\text{ \ }\overset{v}{\Longrightarrow }\text{ }r_{3}^{\prime }=0=\frac{1%
}{\sqrt{2}}\sinh \left( \sqrt{2}\psi _{1}\right) ct+\cosh \left( \sqrt{2}%
\psi _{1}\right) r_{3}.  \label{boost equation 1}
\end{equation}%
Solving equation $\left( \ref{boost equation 1}\right) $ for the velocity $%
v_{3}=\frac{r_{3}}{t_{1}}$%
\begin{eqnarray}
\frac{1}{c}\frac{r_{3}}{t} &=&-\frac{1}{\sqrt{2}}\tanh \left( \sqrt{2}\psi
_{1}\right)  \notag \\
\frac{1}{c}\frac{r_{3}}{2t_{1}} &=&-\frac{1}{\sqrt{2}}\tanh \left( \sqrt{2}%
\psi _{1}\right)  \notag \\
v_{3} &=&-\sqrt{2}c\tanh \left( \sqrt{2}\psi _{1}\right) .
\end{eqnarray}%
In term of $\beta ,$we have 
\begin{equation}
\tanh \left( \sqrt{2}\psi _{1}\right) =-\frac{1}{\sqrt{2}}\frac{v_{3}}{c}=-%
\frac{1}{\sqrt{2}}\beta .
\end{equation}%
Solving the hyperbolic identity for $\cosh \left( \sqrt{2}\psi _{1}\right) ,$
we have 
\begin{eqnarray}
\cosh ^{2}\left( \sqrt{2}\psi _{1}\right) -\sinh ^{2}\left( \sqrt{2}\psi
_{1}\right) &=&1  \notag \\
1-\tanh ^{2}\left( \sqrt{2}\psi _{1}\right) &=&\cosh ^{-2}\left( \sqrt{2}%
\psi _{1}\right) ,  \notag \\
\cosh \left( \sqrt{2}\psi _{1}\right) &=&\frac{1}{\sqrt{1-\tanh ^{2}\left( 
\sqrt{2}\psi _{1}\right) }}  \notag \\
\cosh \left( \sqrt{2}\psi _{1}\right) &=&\frac{1}{\sqrt{1-\frac{1}{2}\left( 
\frac{v_{3}}{c}\right) ^{2}}}=\gamma _{5},
\end{eqnarray}%
where $\gamma _{5}$ is the $5D$ Lorentz factor in $TT$ space. To find the
maximum allowable speed in $TT$ space, we examine the Lorentz factor, $%
\gamma _{5},$ and extract the following constrained inequality%
\begin{equation}
0<1-\frac{1}{2}\left( \frac{v_{3}}{c}\right) ^{2}.  \label{Constraint 8}
\end{equation}%
Solving the inequality $\left( \ref{Constraint 8}\right) $ for $v_{\max },$
we have 
\begin{equation}
v_{\max }<\pm \sqrt{2}c\approx \pm 1.4c,  \label{vmax1}
\end{equation}%
which exceeds the absolute speed $c$ of Minkowski space. The maximum
allowable speed can also be obtained fro equation $\left( \ref{3-vector}%
\right) $%
\begin{eqnarray}
dp^{\widehat{A}} &=&\left( cdt_{1}\left( \psi \right) ,cdt_{2}\left( \psi
\right) ,dr_{3}\left( \psi \right) \right) ,  \notag \\
dp^{\widehat{A}} &=&\left( cdt_{1},cdt_{1},dr_{3}\right) ,
\label{3-vector 1}
\end{eqnarray}%
since the two time-like dimensions are parametrized by the same hyperbolic
function as derived in section 2. The invariant infinitesimal length squared 
$ds^{2}$ can be computed by using the $TT$ inner product on the 3-vector%
\begin{eqnarray}
ds^{2} &=&\left[ TT\right] _{\widehat{A}\widehat{B}}dp^{\widehat{A}}dp^{%
\widehat{B}}=dp^{\widehat{A}}\circledast dp_{\widehat{A}}
\label{infinitesimal length} \\
&=&-c^{2}dt_{1}^{2}-c^{2}dt_{1}^{2}+dr_{3}^{2} \\
ds^{2} &=&-2c^{2}dt_{1}^{2}+dr_{3}^{2}.  \notag
\end{eqnarray}%
To obtain $v_{\max },$ the world-line of the particle must be on the
light-cone, i.e. $ds^{2}=0,$ hence%
\begin{eqnarray}
\left( \frac{dr_{3}}{dt_{1}}\right) ^{2} &=&2c^{2}  \notag \\
\frac{dr_{3}}{dt_{1}} &=&\pm \sqrt{2}c  \notag \\
v_{\max } &=&v_{3}=\pm \sqrt{2}c\approx \pm 1.4c,  \label{vmax3}
\end{eqnarray}%
where $v_{3}=\frac{dr_{3}}{dt_{1}}$ and $v_{\max }=v_{3}<\sqrt{2}c.$

\section{Summary}

The specific forms of the elements, $p^{A}\in M\subset $ $TT$ space and $%
\overset{\cdot }{p}_{A}\in W\subset T_{\overrightarrow{p}}(M)$, where $M$
and $W$ are some open subsets of $TT$ space and its associated tangent space 
$T_{P^{A}}(M)$, were derived via $TT$ metric function. The $TT$ Lorentz \
transformation matrix was fixed or obtained by using the length invariant
condition and the $TT$ metric function. It was shown that the $TT$ Lorentz
transformation matrix transformed a $5D$ vector $p^{A}\in R^{5}$ into $%
\mathbb{C}
^{5}.$ As an application in $TT$ space, aid with the $TT$ Lorentz
transformation matrix, a boost of $v$ along the spatial direction between
the two inertial frames is performed. The resulting boost transformation
yielded a maximum allowable speed, $v_{\max }$ , for particles living in the 
$TT$ space. The maximum speed is shown to be, $v_{\max }<\sqrt{2}c,$ which
exceeds the absolute speed, $c,$ where $c$ is the speed of light in
Minkowski space.

\pagebreak

\end{document}